\documentclass {jetpl}

\twocolumn

\lat

\def\L{{\bf L}}

\title {First-principles investigation of uranium monochalcogenides}

\rtitle {First-principles investigation of uranium monochalcogenides}

\sodtitle {First-principles investigation of uranium monochalcogenides}

\author {A.\,O.\,Shorikov, J.\,E.\, Medvedeva$^{+}$, A.\,I.\, Poteryaev, V.\,V.\, Mazurenko$^{*}$, V.\,I.\,Anisimov,
\thanks {e-mail: alexey.shorikov@gmail.com}}

\rauthor {A.\,O.\,Shorikov, J.\,E.\, Medvedeva, A.\,I.\, Poteryaev et al.}

\sodauthor {A.\,O.\,Shorikov, J.\,E.\, Medvedeva, A.\,I.\, Poteryaev, V.\,V.\, Mazurenko, V.\,I.\,Anisimov}

\address {Institute of Metal Physics, Russian Academy of Sciences, 
620041 Yekaterinburg GSP-170, Russia\\~\\
$^+$Department of Physics, Missouri S\&T, Rolla, Missouri, 65409, USA\\~\\
$^*$Theoretical Physics and Applied Mathematic Department, Urals State Technical University,
620002, Mira st. 19, Yekaterinburg, Russia}

\dates {\today}{*}

\abstract {We present first-principles investigation of the electronic structure and
magnetic properties of uranium monochalcogenides:  US,  USe,  UTe. 
The calculations were performed by using recently developed LDA+U+SO method in which both
Coulomb and spin-orbit interactions have been taken into account in rotationally invariant form.
We discuss the problem of choice of the Coulomb interaction value.
The calculated [111] easy axes agree with those experimentally observed.
The electronic configuration 5$f^3$ was found for all uranium compounds under investigation.}


\begin {document}

\maketitle

Local Spin Density Approximation (LSDA)~\cite{KohnSham} based on the density functional theory is
a widely used method for electronic and magnetic structure investigations of modern materials.
Despite the success in description of wide band materials, LSDA fails when applied to describe transition
metal or rare earth metal compounds. For example, it gives a metallic ground state in case of
3$d$ insulators (such as CuO, CoO, FeO) or underestimates an energy gap and
local magnetic moments values for NiO~\cite{TOWK84}. Another important drawback of LSDA is an
underestimation of the orbital moment value {\bf L}.
LSDA calculations with the spin-orbit (SO) coupling taken into account yield
the orbital moment value about two times smaller then its experimentally observed counterpart~\cite{Singh,Chen}.
While {\bf L} is small for transition metal compounds and one can neglect it,
this is not the case for 4$f$- and 5$f$-metal systems where the value of the orbital moment is
larger than the spin one.

The problem of the underestimation of \L ~ comes from the orbital independent nature of the LSDA
potential and can be solved by using LDA+U+SO approach, where
spin and orbital dependent on-site potential is provided~\cite{shorikov}.
It results in orbital polarization increasing and, hence,
it increases the value of the orbital moment and the magnetic anisotropy energy~\cite{Shick}.
The results of previous theoretical~\cite{Tatsuya} and experimental~\cite{Busch} studies demonstrate
that the orbital polarization mechanism plays a crucial role in 5$f$ actinide systems, such as US, UTe and USe.



In this paper we report the results of first-principle LDA+U+SO investigations of uranium
monochalcogenides: US,  USe and UTe.
Choice of the screened Coulomb interaction is discussed and the obtained results are
compared with the LDA+SO and experimentally observed one.

The investigated uranium monochalcogenides have the NaCl-type crystal structure and
are ferromagnets with Curie temperature 178~K, 160~K and 102~K for US, USe, UTe respectively~\cite{Fournier}.
Despite of simple and high symmetric crystal structure ($Fm\bar3m$), the easy axes are arranged
along the diagonal of the cubic cell [111]~\cite{Twillick,Busch,Kraft}.
In our calculations we use the tight-binding linear muffin-tin orbital approach
in atomic sphere approximation (Stuttgart LMTO47 code)~\cite{Andersen75}
with conventional local-density approximation and take into account
the on-site Coulomb and spin-orbit interactions (LDA+U+SO)~\cite{shorikov}.
The LMTO basis set contains the following states: U(7$s$,6$p$,6$d$), S(3$s$,3$p$,3$d$),
Se(4$s$,4$p$,4$d$) and Te(5$s$,5$p$,5$d$,5$f$).
The Brillouin zone integration has been performed on the 8x8x8 grid.

%
%

The value of screened Coulomb interaction, $U$, and Hund's exchange, $J_H$, are cornerstones
of the LDA+U+SO method. While determination of former is a complicated issue and depends
strongly from screening in solids, the later is hardly changed from its atomic (ionic) value.
The constrain LDA method described in Ref.~\cite{Gunnarsson89} gives
the value $J_{H}$ = 0.48~eV for all compounds and
it is in good agreement with estimated in Ref.~\cite{shorikov}.
This value of Hund's exchange will be used through the paper.

It is well-known that the $U$ value depends strongly from number of screening channels taken into account.
For a free ion the Coulomb parameter, $U$, is about 20~eV, while its value in the solid
varies from 4~eV to 10~eV for 3$d$-metals.
In actinides, the value of the on-site Coulomb interaction becomes smaller
since 5$f$ states are more expanded in real space than 3$d$ one.
Several features that can be calculated using band structure methods, e.g.
magnitude of local magnetic moment, density of states at the Fermi energy for metals or
energy gap for insulator depends strongly on the Coulomb parameter.
Therefore, the first important task of investigation is to define $U$ in reliable way.
In this paper we demonstrate different approaches to calculate the value of screened Coulomb repulsion.

\begin{figure}
  \vspace{6mm}
  \centering
  \includegraphics[width=0.45\textwidth]{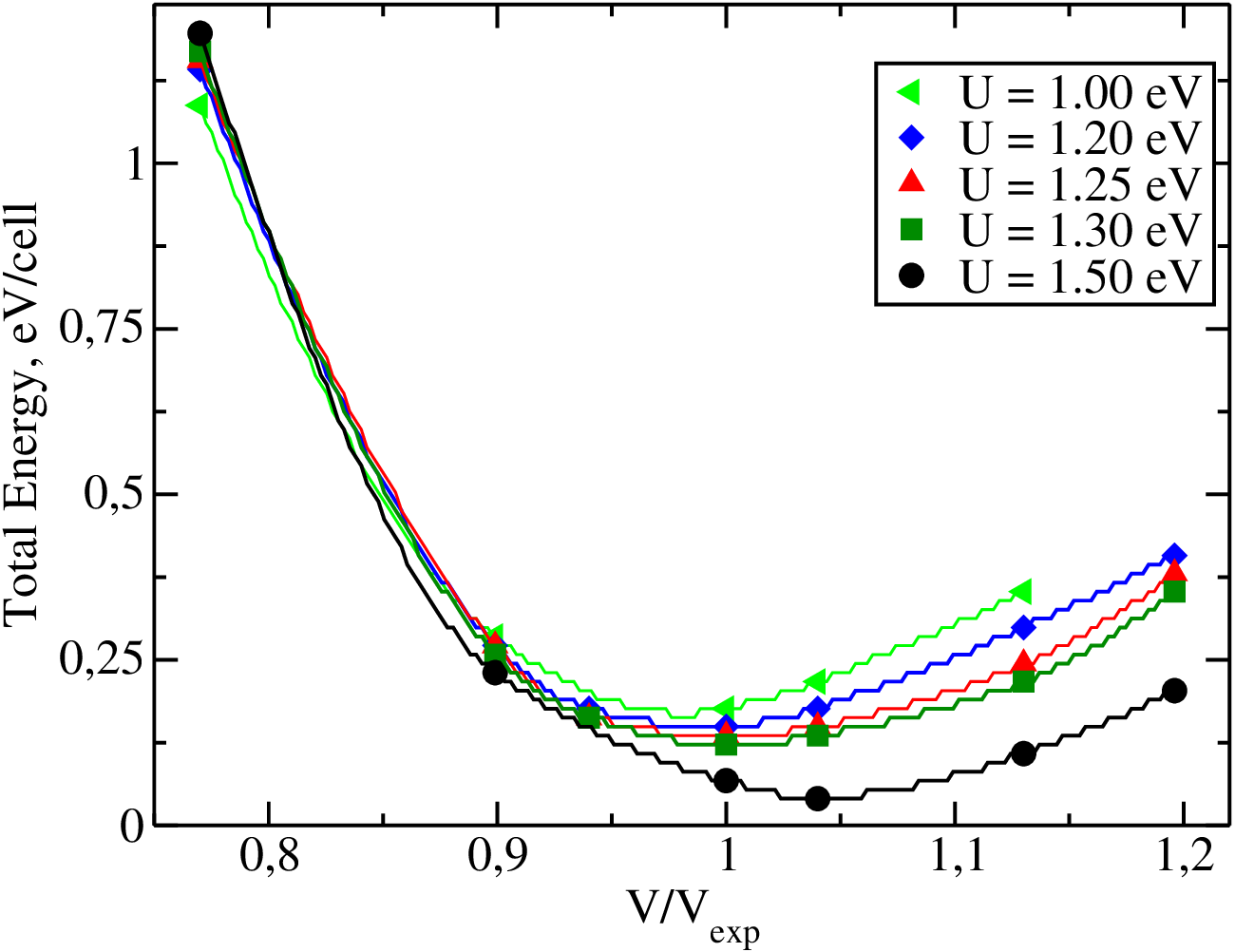}
  \caption{Total energy vs. cell volume for different $U$ values obtained in LDA+U+SO calculations for US.}
  \label{eq_vol_US}
\end{figure}

One way is to adapt the Coulomb interaction value to have certain
calculated and experimentally measured physical quantities coinciding.
For example, T.~Shishidou {\it et al.}~\cite{Tatsuya} have shown using a Hartree-Fock approximation
with $U$ as a free parameter that calculated and experimental values of the local magnetic moments and
the direction of the easy axes agree well for $U$=0.76~eV.
In this work we fit an equilibrium volume of US by varying $U$-parameter.
The result is shown on the Fig.~\ref{eq_vol_US}.
The calculated equilibrium volume is close to the experimental one
at the value of $U$=1.25~eV and it will be used
for the uranium compounds under investigation.

\begin{figure}
  \vspace{6mm}
  \centering
  \includegraphics[width=0.45\textwidth]{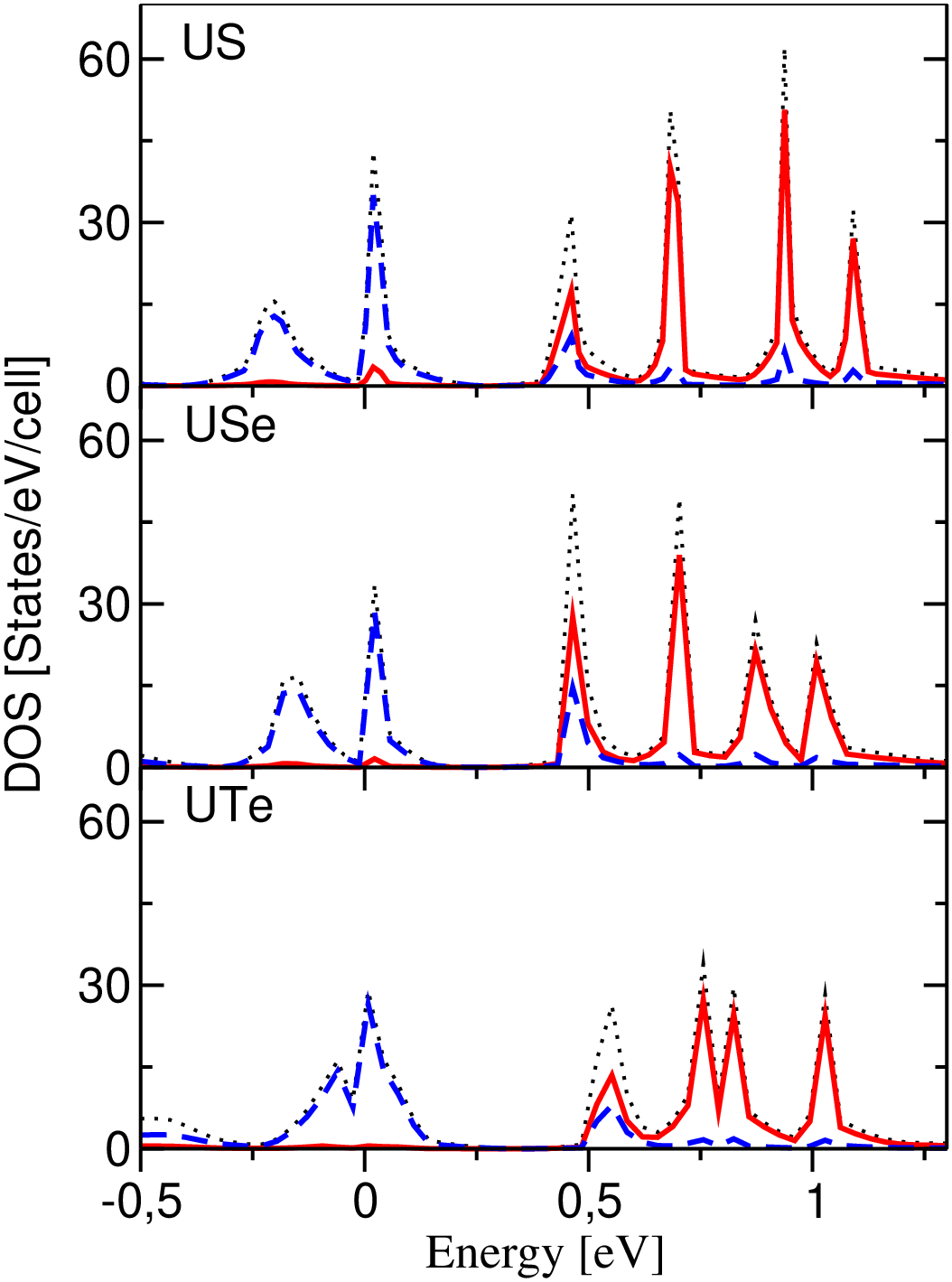}
  \caption{LDA+SO partial DOS for uranium monochalcogenides.
           $j$=5/2 is represented by dashed blue line,
           $j$=7/2 is represented by solid red line and its sum is a dotted black curve.}
  \label{fig:p_dos}
\end{figure}

Another way to calculate the value of screened Coulomb interaction
from first principles is the supercell procedure~\cite{Gunnarsson89}
that takes into account $s$-, $p$- and $d-$ screening channels.
In the framework of this procedure, the calculated Coulomb repulsion is equal to 3.6~eV for US.
This value is larger than reported by Shishidou~\cite{Tatsuya} and have found in this work.
To understand this discrepancy we will analyze the LDA+SO densities of states presented
in Fig.~\ref{fig:p_dos}.
One can clearly see that $j$=5/2 and $j$=7/2 subbands are well separated due to
a strong spin-orbit interaction (the value of spin-orbit coupling $\lambda$=??~eV)
and, hence, for the investigated uranium monochalcogenides
one more screening channel should be taken into account.
The additional screening channel of conducting $j$=7/2 states
leads to much smaller $U$ values which are
0.82~eV, 0.98~eV and 0.75~eV for US, USe and UTe, respectively.
These values are closer to one used in Ref.~\cite{Tatsuya}.

We have performed two series of calculation {\it i)} with the value of Coulomb interaction
obtained by fit of equilibrium volume, $U$=1.25~eV, for all three compounds and
{\it ii)} with three different $U$ values calculated in supercell procedure which
takes into account 7/2-5/2 screening.

We emphasize here that in all our calculations we have started
with initial directions of spin and orbital moments coincided with the crystallographic $c$-axes.
The resulting self-consisted {\bf S} and {\bf L} vectors were found to be anti-parallel to each other and
aligned along [111] direction. It coincides with easy axes experimentally observed
for all compounds~\cite{Twillick,Busch,Kraft}.

\begin{figure}
  \vspace{6mm}
  \centering
  \includegraphics[width=0.45\textwidth]{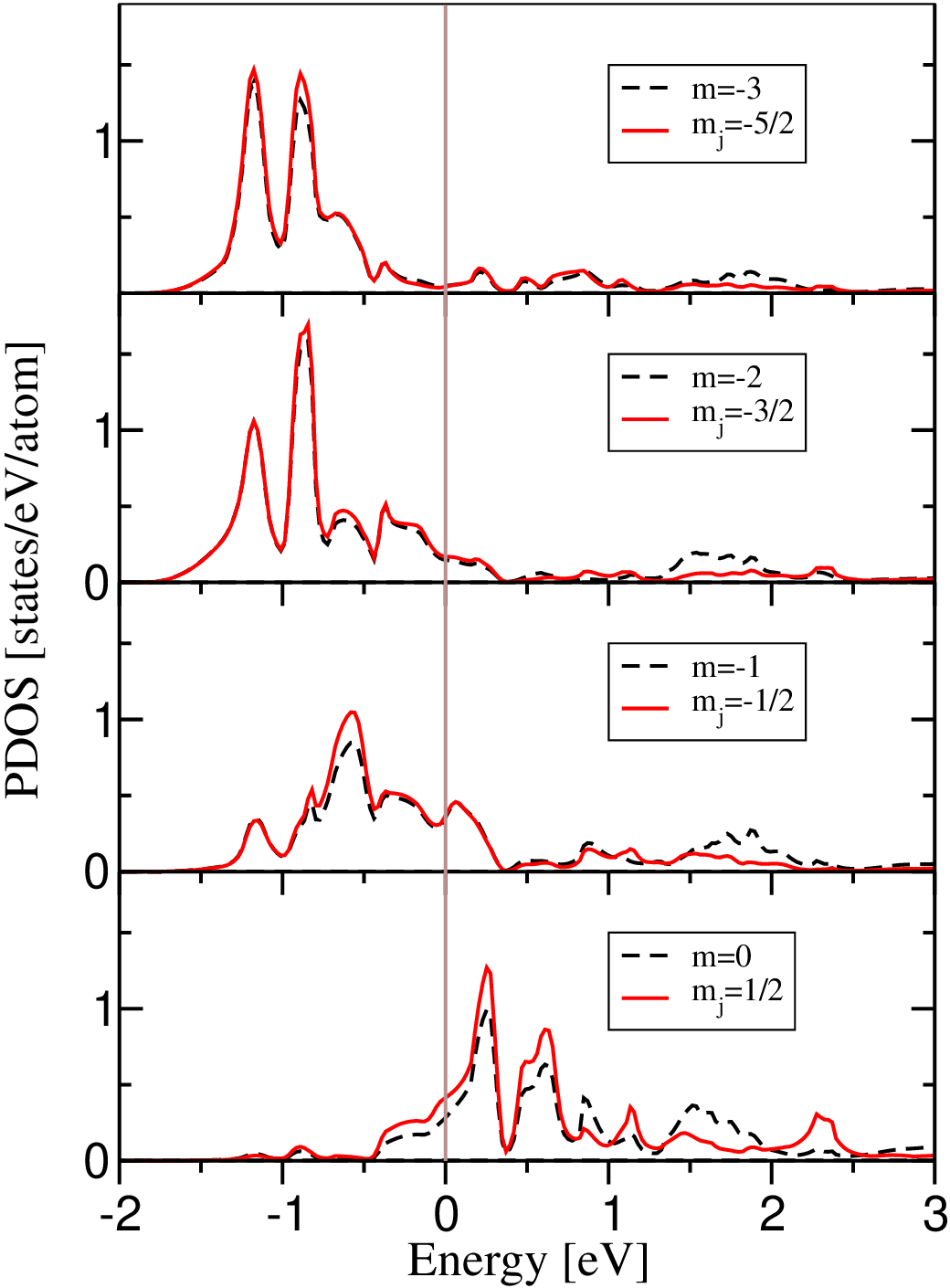}
  \caption{LDA+U+SO partial DOS for US. Four most occupied orbitals shown only.
           Density of states in Russell-Saunders ($LS$) representation
           is shown by dashed black line. Red solid line shows PDOS in $jj$ coupling
           scheme.}
  \label{dos_jmj_all}
\end{figure}

In order to analyze different contributions of the 5$f$ states to the resulting densities of states
we have chosen a new local coordinate system with $z$-axis directed
along total moment {\bf J}.
The partial density of states for 4 orbitals with maximal occupation numbers
in both $jj$- and $LS$-representations are shown in the Fig.~\ref{dos_jmj_all}.
Based on the obtained densities of states we can conclude
that there are three occupied orbitals and one partially occupied orbital
which forms the peak on the Fermi level.
The similarity of PDOS shape in both $jj$- and $LS$-schemes confirms
the applicability of intermediate coupling scheme.
Another argument to confirm the intermediate character of coupling is
an occupation matrix which has a large off-diagonal elements
in either $jj$ or $LS$ coupling schemes. The magnitude of the largest off-diagonal
element in both schemes is just {\bf ??} per cents smaller than diagonal one.
This result can be explained as a strong competition of a strong spin-orbit interaction
and intra-atomic exchange interaction. Indeed, the values of spin-orbit coupling,
$\lambda$=??~eV, is comparable with the Hund's exchange, $J_H$=0.48~eV.
Therefore, in the situation where the Hund's exchange, voting for the Russell-Saunders
representation, cannot beat the spin-orbit interaction, favoring $jj$ representation,
the intermediate coupling scheme is more desirable for all compounds under consideration.


Experimentally observed values of the 5$f$ magnetic moments are much smaller
than expected for a free ion with $f^3$ configuration (see Table~\ref{mom_2}).
We have calculated the effective magnetic moments for US, USe, UTe using the following expression:
$\mu_{eff}^2 = g \mu_{B} J(J+1)$ where Land\'e factors for $f^3$ configuration are
$g_{LS}$=0.73 and $g_{jj}$=0.86 for $LS$ and $jj$-couplings, respectively.
The comparison of the calculated and experimental magnetic moments is presented in Tables~\ref{mom_2}
and~\ref{mom_3}.
Theoretical moments seem to be overestimated but howbeit agree reasonably well with the experimental data.
This overestimation is a result of the large on-site Coulomb interaction, $U$=1.25~eV.
The results of second series of LDA+U+SO calculation with $U$=0.82~eV, 0.98~eV and 0.75~eV
for US, USe and UTe, respectively, are presented in Table~\ref{mom_3}.
One can clearly see that there is a good agreement between theoretical and experimental values.

\begin{table}
  \centering
  \begin{tabular}{l|l|l|l|l|l}
    & J    & $\mu^{calc}_{jj}$ & $\mu^{calc}_{LS}$ & $\mu_{eff}$~\cite{Busch78} & $\mu_{neut}$~\cite{Wedgwood72} \\
   \hline
US  & 3.18 &   3.14            &   2.66            &   2.2                      & 1.7$\pm$0.03   \\
USe & 3.38 &   3.31            &   2.81            &   2.5                      & 2.0$\pm$0.1    \\
UTe & 3.51 &   3.42            &   2.91            &   2.8                      & 2.2$\pm$0.1    \\
  \end{tabular}
  \caption{Comparison of LDA+U+SO calculation results obtained for $U$=1.25~eV and
           experimental data ($g_{LS}$=0.73, $g_{jj}$=0.86).}
  \label{mom_2}
\end{table}

\begin{table}
  \centering
  \begin{tabular}{l|l|l|l|l|l}
    &  J   & $\mu^{calc}_{jj}$ & $\mu^{calc}_{LS}$ & $\mu_{eff}$~\cite{Busch78} & $\mu_{neut}$~\cite{Wedgwood72} \\
  \hline
US  & 2.78 &    2.79           &   2.37            &    2.2                     &    1.7$\pm$0.03   \\
USe & 3.18 &    3.14           &   2.66            &    2.5                     &    2.0$\pm$0.1    \\
UTe & 3.30 &    3.24           &   2.75            &    2.8                     &    2.2$\pm$0.1    \\
  \end{tabular}
  \caption{Comparison of LDA+U+SO calculation results (Coulomb interaction values are 0.82~eV,
           0.98~eV and 0.75~eV for US, USe and UTe, respectively) and experimental data
           ($g_{LS}$=0.73, $g_{jj}$=0.86).}
  \label{mom_3}
\end{table}

To conclude, we have performed first-principles investigations of uranium monochalcogenides: US, USe and UTe.
The choice of Coulomb parameter $U$ has been discussed. We have demonstrated that supercell procedure
with additional screening channels produces the value of screened Coulomb interaction which
gives the values of effective magnetic moments in better agreement with the experimental data.
It can be traced to the fact that this is an electronic degrees of freedom based method while
fitting of $U$ to have the calculated and experimental volumes coinciding should better
describe phonon properties of solids.
We have analyzed LDA+U+SO results in two different coupling basis of $LS$ and $jj$ types.
It was found that for the studied uranium compounds the intermediate coupling scheme is more preferable.
Based on the obtained occupation matrices and partial densities of states we have concluded
about 5$f^3$ electronic configuration for the uranium compounds.

Authors acknowledge M.A. Korotin for fruitful discussion.
This work was supported by grants RFBR 04-02-16096, RFBR 09-02-00431a, RFBR-GFEN 03-02-39024,
the grant program of President of Russian Federation MK-1162.2009.2,
the Russian Federal Agency of Science and Innovation N 02.740.11.0217,
the scientific program ``Development of scientific potential of universities'' N 2.1.1/779,
NWO-047.016.005, grant UB and SB of RAS $\mathcal N^{\circ \!\!\!\!\_~}$22.


\begin{thebibliography}{99}
\bibitem{KohnSham}
W. Kohn and L.J. Sham, Phys. Rev. A {\bf 140}, 4, 1133, (1965).
\bibitem{TOWK84}
K. Terakura,  T. Oguchi, A.R. Williams and J. K\"ubler,  Phys.
  Rev. B.   {\bf 30}, 4734--4747, (1984).
\bibitem{Singh}
M. Singh, J. Callaway and C.S. Wang,  Phys Rev. B {\bf 14}, 3, 1214, (1976).
\bibitem{Chen}
C.T. Chen, Y.U. Idzera, H.-J. Lin, N.V. Smith, G. Meigs, E. Chaban, G.H. Ho, E. Pellegrin and F. Sette, Phys Rev. Lett. {\bf 75}, 1, 152, (1995).
\bibitem{Fournier}
J.-M. Fournier and R. Troc,  Handbook on the Physics and Chemistry of the
  Actinides. Edited by A.J. Freeman and G.H. Lande,  North-Holland, Amsterdam. 1985.
\bibitem{Twillick}
 D.L. Twillick and P. de~V.~du Plessis , J. Magn. Magn. Mater.  {\bf 3}, (4), 329, (1976).
\bibitem{Busch}
G. Busch, O. Vogt, A. Delpalme and G.H. Lander, J. Phys. C {\bf12}, (7), 1391 (1979).
\bibitem{Kraft}
 T. Kraft, P.M. Oppener, V.N. Antonov  and H. Eschrig, Phys. Rev. B {\bf 52}, 3561--3570, (1995).
\bibitem{Lander90}
G.H. Lander, S.S. Brooks, B. Lebech, P.J. Brown, O. Vogt and K. Mattenberg, Appl. Phys. Lett.  {\bf 57}, 10, 989, (1990).
\bibitem{shorikov}
A.O. Shorikov, A.V. Lukoyanov, M.A. Korotin and V.I. Anisimov, Phys Rev. B {\bf 72}, 024458, 2005.
\bibitem{Shick}
A.B. Shick and A.I. Lichtenstein, J. Phys.: Cond. Matter. {\bf 20}, 015002, 2008.
\bibitem{Andersen75}
 O.K. Andersen, Phys. Rev. B {\bf 12}, 8, 3060--3083 (1975).
\bibitem{ldau}
V.I. Anisimov, F. Aryasetiawan and A.I. Lichtenstein, J. Phys.: Condens. Matter.  {\bf 9}, 767--808, (1997).
\bibitem{Gunnarsson89}
O. Gunnarsson, O.K. Andersen, O. Jepsen  and J. Zaanen, Phys. Rev. B {\bf 39}, 3, 1708--1722, (1989).
\bibitem{Tatsuya}
T. Shishidou and T. Oguchi, Phys Rev. B  {\bf 59}, 10, 6813--6823, (1999).
\bibitem{Busch78}
G. Busch  and  O. Vogt, J. Less-Common Met. {\bf 62}, 335, (1978).
\bibitem{Wedgwood72}
 F.A. Wedgwood, J. Phys. C {\bf 5}, 17, 2427,  (1972).
\end{thebibliography}
\end{document}